\documentclass{article}
\usepackage{times}
\usepackage{graphicx}
\usepackage{amssymb}
\usepackage{amsmath}
\usepackage{xcolor}
\usepackage{latexsym}
\usepackage{verbatim}
\usepackage{fancyhdr}
\usepackage{lineno}
\usepackage{booktabs}
\DeclareUnicodeCharacter{0301}{\'{e}}

\usepackage[style=nature]{biblatex}

\title{Scale-dependent irreversibility in living matter}

\author
{Tzer Han Tan$^{1,2\dagger}$, Garrett A. Watson$^{1\dagger}$, Yu-Chen Chao$^{1\dagger}$, Junang Li$^{1}$, Todd R. Gingrich$^{3}$, \\
	Jordan M. Horowitz$^{4,5,6\ast}$, Nikta Fakhri$^{1\ast}$\\
\\
\normalsize{$^{1}$Department of Physics, Massachusetts Institute of Technology, Cambridge, MA 02139, USA}\\
\normalsize{$^{2}$Center for Systems Biology Dresden, Max Planck Institute of Molecular Cell Biology and Genetics,}\\
\normalsize{Dresden 01307, Germany (\emph{current address})}\\
\normalsize{$^{3}$Department of Chemistry, Northwestern University, Evanston, IL 60208, USA}\\
\normalsize{$^{4}$Department of Biophysics, University of Michigan, Ann Arbor, Michigan, 48109, USA}\\  
\normalsize{$^{5}$Center for the Study of Complex Systems, University of Michigan, Ann Arbor, Michigan, 48109, USA}\\
\normalsize{$^{6}$Department of Physics, University of Michigan, Ann Arbor, Michigan, 48109, USA}\\
\\
\normalsize{$^\ast$Corresponding authors.}\\
\normalsize{$^\dagger$These authors contributed equally.}
}

\date{}

\begin{document}

\baselineskip24pt

\maketitle

\section*{Abstract}
A defining feature of living matter is the ability to harness energy to self-organize multiscale structures whose functions are facilitated by irreversible nonequilibrium dynamics~\cite{needleman2017active}. While progress has been made in elucidating the underlying principles, what remains unclear is the role that thermodynamics plays in shaping these structures and their ensuing functions. Here, we unravel how a fundamental thermodynamic connection between the physical energy dissipation sustaining a nonequilibrium system and a measure of statistical irreversibility (arrow of time), can provide quantitative insight into the mechanisms of nonequilibrium activity across scales~\cite{Parrondo2009}.
Specifically, we introduce a multiscale irreversibility metric and demonstrate how it can be used to extract model-independent estimates of dissipative timescales. Using this metric, we measure the dissipation timescale of a multiscale cellular structure -- the actomyosin cortex -- and further observe that the irreversibility metric maintains a monotonic relationship with the underlying biological nonequilibrium activity. 
Additionally, the irreversibility metric can detect shifts in the dissipative timescales when we induce spatiotemporal patterns of biochemical signaling proteins upstream of actomyosin activation.
Our experimental measurements are complemented by a theoretical analysis of a generic class of nonequilibrium dynamics, elucidating how dissipative timescales manifest in multiscale irreversibility.

\section*{Main text}
Living systems operate far from equilibrium, constantly consuming energy. This energy is harnessed at the molecular-scale through ATP hydrolysis and dissipated on larger spatiotemporal scales spanning the entire cell. Canonical examples of these cellular structures are the actomyosin cortex~\cite{chugh2018actin,tan2018self} and mitotic spindle~\cite{sawin1992mitotic,brugues2014physical}, which are required for vital functions such as cell division. These living systems are, in fact, examples of the broader class of active matter, where large scale rearrangements and functions are driven by energy-consuming microscopic processes, such as mechanochemical reactions~\cite{needleman2017active}. 
The inherent multiscale nature of such systems makes it challenging to detect the nonequilibrium activity, let alone unravel its characteristic features~\cite{fakhri2014high,guo2014probing}.

While various methods have been developed to detect the existence of nonequilibrium activity, they often fail to reveal any mechanistic insight.
For instance, a clear signature of a nonequilibrium process is directed motion or flux.
This observation has led to a handful of methods to identify the existence of such fluxes and to extract from their fluctuations estimates (or bounds) on the nonequilibrium dissipation, typically in individual mesoscopic systems~\cite{Zia2016,Battle2016,gladrow2016broken,martinez2019inferring, Steinberg1986,gnesotto2020learning,kim2020learning,otsubo2020estimating,manikandan2020inferring,otsubo2020estimating2}.
However, there is no clear method available to extract characteristic features of the nonequilibrium activity from the estimates of these fluxes.
On the other hand, departures from the fluctuation-dissipation theorem (FDT) are inherently connected to nonequilibrium activity~\cite{Harada2005,Nardini2017,prost2009generalized} in a way that could reveal more detailed information about the timescales of nonequilibrium fluctuations; however, 
such analyses are typically focused on identifying structure in a time-dependent (or frequency-dependent) effective temperature whose interpretation usually requires prior information about the mechanisms~\cite{martin2001comparison,mizuno2007nonequilibrium,BenIsaac2011, Dieterich2015, Crisanti2003, Cugliandolo2011, Fodor2016}. Specifically for mesoscopic diffusive systems, the deviations of the velocity FDT can be used to estimate the dissipation using the Harada-Sasa equality~\cite{Harada2005}, an approach that has been successfully utilized to estimate the energetic efficiency of molecular motors~\cite{toyabe2010nonequilibrium,ariga2018nonequilibrium} and to probe the statistics of active forces~\cite{bohec2013probing}.
Even still, the FDT approach requires active perturbations in order to measure response properties, which in practice can be challenging.
Thus, the development of new frameworks that can identify the scales of energy dissipation is crucial for a mechanistic understanding of nonequilibrium processes in complex biological and active matter systems.

\subsection*{Irreversibility measures the arrow of time}
Any nonequilibrium process is accompanied by the irrecoverable loss of energy to its environment, often in the form of heat, which commensurately produces entropy in the environment. The unidirectionality of this flow implies a direction to time’s arrow. In other words, the process is inherently irreversible~\cite{Maes2002,Maes2003,Jarzynski2006a,Kawai2007,Parrondo2009}.
Here, we show that by measuring irreversibility one can reveal the temporal structure of nonequilibrium activity from experimental time series data collected in a living system -- the actomyosin cortex of a starfish oocyte. 
Thereby, we discover that recent developments in nonequilibrium thermodynamics that were previously thought only able to confirm or deny the existence of nonequilibrium~\cite{Roldan2010,Roldan2012,roldan2018arrow} can, in fact, be adapted to unravel timescale information about dissipative processes. 
Nonequilibrium thermodynamics offers a quantitative connection between dissipation and the statistical irreversibility (time-reversal asymmetry) of the dynamics.
For a nonequilibrium steady state, the relative entropy or Kullback-Leibler divergence (KLD) between the probability ${\mathcal P}(x_1,\dots,x_n)$ to observe a stationary series of measurements $(x_1,\dots,x_n)$ of a time-symmetric observable and the probability to observe the reverse sequence ${\mathcal P}(x_n,\dots,x_1)$ bounds the dissipation (entropy production) as~\cite{Maes2003,Jarzynski2006a,Parrondo2009,gaveau2014relative,gaveau2014dissipation}
\begin{equation}\label{eq:KLD}
\Sigma \ge \sigma=\lim_{n\to\infty}\sigma_n \equiv  \lim_{n\to\infty}\frac{k_{\rm B}}{n\tau}D[{\mathcal P}(x_1,\dots, x_n)||{\mathcal P}(x_n,\dots,x_1)],
\end{equation}
where $\tau$ is the sampling interval and $k_{\rm B}$ is Boltzmann's constant.
Generically, the entropy production $\Sigma$ can be related to the energy dissipation once one identifies the nature of the thermodynamic reservoirs in the environment with which the system exchanges energy: For example, chemical reactions occurring at rate $r$ driven by a chemical potential difference $\Delta\mu$ produce entropy equal to the chemical work $\Sigma=r\Delta\mu$. The KLD, $D(p||q)=\int dx\ p(x)\ln p(x)/q(x)$, is an information-theoretic measure of distinguishability that generically decreases under coarse graining~\cite{Cover}, with equality when every nonequilibrium degree of freedom is included. As such, $\sigma$ measures the observed statistical distinguishability between the forward and reverse senses of time, making it a quantitative measure of irreversibility.
With this in mind, we call $\sigma$ the (observed) irreversibility.
In the absence of detailed knowledge of the system, Eq. (1) has been developed into a robust experimental and computational tool to detect nonequilibrium activity even in the absence of observable flows~\cite{Kawai2007,Parrondo2009,Gomez-Marin2008a,Gomez-Marin2008b,Horowitz2009b,Roldan2010,Roldan2012,Muy2013,Tusch2014}.
Here, we exploit the flexibility in the choice of observable in Eq. (1) to probe multiple timescales of irreversibility by repeatedly estimating $\sigma$ for progressively more coarse-grained time-series data.
Coarse graining in time allows us to smear over dynamic processes with fast characteristic timescales, thereby removing their influence on the observed irreversibility.
Then by tracking the variation of irreversibility with the coarseness of the data, we can unravel timescales important to dissipation.
Such a coarse-graining approach to time series analysis has previously been successfully employed as a basis for the multiscale sample entropy method for complexity assessment~\cite{Costa2002}.

\subsection*{Actomyosin cortex: a multiscale dissipative structure}
\begin{figure}[ht]
	\includegraphics[width=0.9\textwidth]{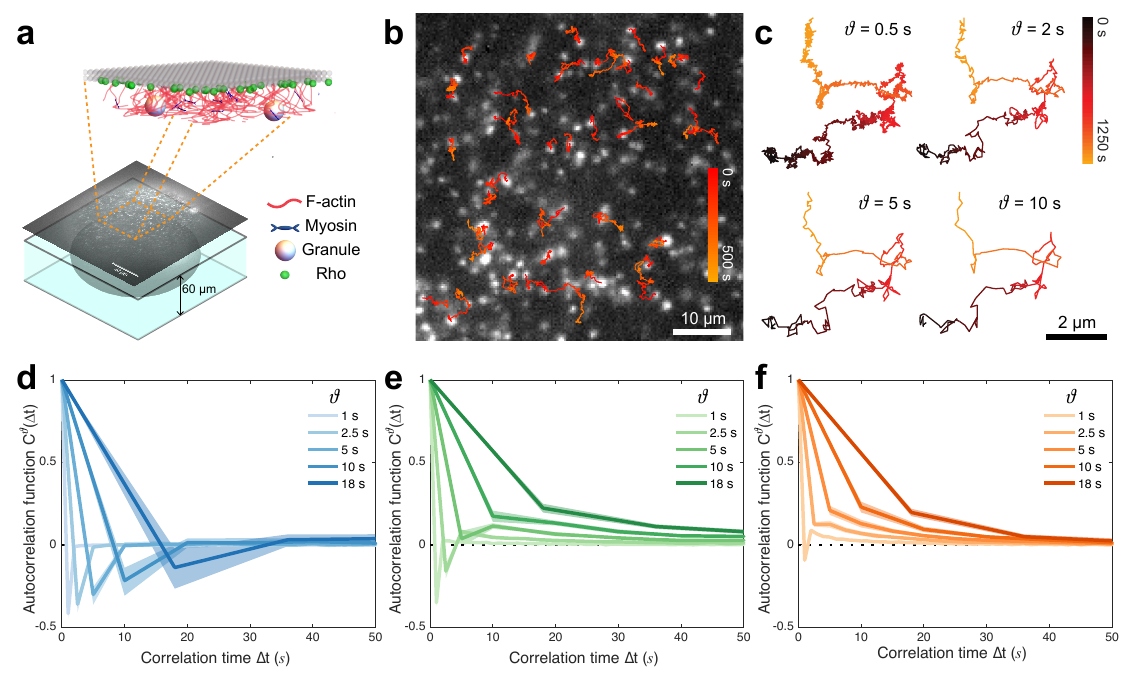}
	\caption{{\bf Positional fluctuations and position-increment autocorrelation function of endogenous cortical granules probe local force fluctuations of the actomyosin cortex.} ({\bf a}) Cortical cross-section of a flattened starfish oocyte imaged using widefield near-infrared fluorescence microscopy.  Puncta represent individual granules. Schematic shows the cortical granules (yellow) embedded in the actin network (red) activated by myosin motor minifilaments (blue). ({\bf b}) Example position trajectories of embedded cortical granules color coded with time. ({\bf c}) 2D coarse-grained trajectories at different lag time $\vartheta$ color coded by time. ({\bf d-f}) Comparison of position-increment autocorrelation function $C^\vartheta(\Delta t)$ with various lag times $\vartheta = 1,2.5,5,10,18\ s$ as a function of correlation time $\Delta t$ for an oocyte with (d) low activity (ATP depletion), (e) medium activity (wild type), and (f) high activity (myosin over expression) displays  an activity-dependent cross-over with coarse-graining lag time. Notice how the observed short-time elastic response characterized by a sudden and rapid drop disappears at lower lag times for higher activity treatments.} 
	\label{fig1}
\end{figure}

The actomyosin cortex is a thin, cross-linked network of filamentous actin and myosin motors that is anchored to the cell membrane~\cite{chugh2018actin}. The cortex of an immature starfish oocyte is in a nonequilibrium steady state that persists over very long times ($> 24\ hours$ post extraction). The nonequilibrium processes in the cortex, in particular the force fluctuations generated by the activity of myosin motors, are powered by constant ATP consumption at the molecular scale. We varied the amount of the nonequilibrium activity in the actomyosin cortex via ATP depletion treatments (low activity steady state) or over-expression of myosin motors (high activity steady state)~\cite{bischof2017cdk1}. To probe the local force fluctuations, we took advantage of endogenous fluorescent cortical granules ($\sim1\ \mu m$ in size) that are embedded in the cortex (Figs.~\ref{fig1}a-c 
).
With high photostability and modest photobleaching for up to an hour, they enabled long duration ($> 1\ hour$), high-speed microscopy ($> 10\ Hz$). To verify that the cortical granules are embedded in the actin cortex, we depolymerized actin using cytochalasin D and observed that these granules dissociated from the cortex. 
We used a single molecule tracking algorithm~\cite{jaqaman2008robust} to obtain the centroid positions $r(t)$ of the cortical granules (Fig.~\ref{fig1}b) as a function of time $t$, sampled every $\tau$ seconds. We verified that the actomyosin network is spatially homogeneous over the course of our observations (Fig. S1) and stitched particle trajectories from individual granules together into one long trajectory of length $T$ seconds for subsequent analysis.

To estimate the irreversibility $\sigma$, we require data obtained from a stationary steady-state process.
Here, the underlying dynamics of the actomyosin network is stationary, which leads to stationary local forces fluctuations acting on the cortical granules.
Considering that the diffusive dynamics of the cortical granules are overdamped \cite{fodor2015activity}, we can interpret changes in position as  readouts of these stationary local force fluctuations of the underlying nonequilibrium actomyosin network. 
We estimated these position increments $\Delta r^\vartheta(t)=r(t+\vartheta)-r(t)$ using an adjustable lag time $\vartheta$, with larger $\vartheta$ corresponding to more coarse-grained data (Fig.~\ref{fig1}c and S2).
(Stationarity of the time-series was confirmed by a unit root test.)
Thus by varying $\vartheta$, we generate a collection of observables whose estimated KLD $\sigma(\vartheta)$ captures the irreversibility on scales larger than $\vartheta$, scales that have not been smeared out by coarse graining. It is important to note that such a measure is not sensitive to the frequency of sampling (as long as the sampling is faster than the rate of dissipation), as less-frequent sampling manifests in the formalism as simply an additional form of coarse graining.

As a first step in understanding the underlying dynamical processes in the actomyosin cortex, we analyzed the normalized time-lagged position-increment autocorrelation function $C^\vartheta(\Delta t)=\langle \Delta r^\vartheta(t+\Delta t)\Delta r^\vartheta(t)\rangle/\langle \Delta r^\vartheta(t)^2\rangle$ (Fig.~\ref{fig1}d-f). In the low activity steady state (Fig.~\ref{fig1}d), the position increments are delta-correlated at $\Delta t=0$ followed by a negative correlation for small $\Delta t$ with diminishing magnitude for longer lag times $\vartheta$.
This is consistent with the dynamics of the cortical granules being driven primarily by thermal fluctuations in a locally elastic network on short times, as evidenced by the mean squared displacement curve (Fig. S3)~\cite{fakhri2014high,guo2014probing}: small displacements are immediately followed by a displacement that returns the particle to its original position, resulting in oppositely oriented increments.
On longer times, nonequilibrium activity overcomes the elasticity of the network.
This effect is stronger in medium and high activity steady states (Figs.~\ref{fig1}e,f), where the short time equilibrium elastic response (negative dip in the autocorrelation function) is smeared out at smaller lag times: $\vartheta^*\approx 2.5-10\ s$ for the medium activity steady state and $\vartheta^*\approx 1-2.5\ s$ for the high activity steady state.
One might suspect that these times roughly correlate with the timescale of a nonequilibrium mechanism, since they are sensitive to the level of activity.
In the following, we corroborate this conclusion by demonstrating the emergence of this timescale in an analysis of the multiscale irreversibility, which by Eq. (1) offers us a model-independent verification that these timescales are indeed related to the dissipation.

\subsection*{Irreversibility reveals nonequilibrium timescales}
Using the position-increment time-series data, we interrogate the irreversibility in the network's force fluctuations (which we take to be even under time reversal) by estimating the KLD $\sigma(\vartheta)$ between the position increment trajectory $\Delta r^\vartheta(t)$ and its reverse $\Delta r^\vartheta(T-t)$ for various values of lag time $\vartheta$. We used a $k$-nearest-neighbor method  to estimate the KLD~\cite{Wang2009} as detailed in the SI Sec.~1.6~(Fig. S2, S4-5).
For lag times $\vartheta  \geq 8\ s$, we verified that our KLD estimates plateaued with number of samples prior to reaching our sample size of $9.1\times 10^5$ data points.
For lag times $\vartheta < 8\ s$, thermal fluctuations are dominant, and despite increasing the sample size to $3.1\times 10^6$ data points, the convergence is still not satisfactory.

\begin{figure*}[ht]
	\includegraphics[width=\textwidth]{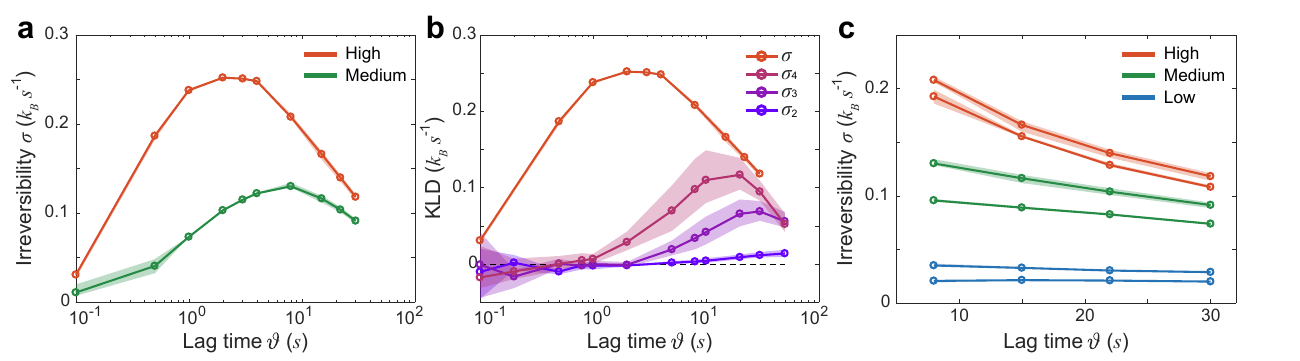}
	\caption{{\bf Time irreversibility reveals the intrinsic timescale of nonequilibrium activity and correctly rank orders nonequilibrium activity above the dissipation timescale.} ({\bf a}) Estimated irreversibility $\sigma$ for the high ($N\approx10\times10^6$ for $\vartheta<5\ s$, $N\approx1\times10^6$ for $\vartheta>5\ s$) and medium ($N\approx5\times10^6$ for $\vartheta<5\ s$, $N\approx1\times10^6$ for $\vartheta>5\ s$) activity treatment as a function of lag time $\vartheta$. The shaded region for low lag time ($\vartheta<5\ s$) of the high and medium activity curves shows the range of 5 bootstraps statistics with $N\approx2.5\times10^6$ and $N\approx1.5\times10^6$ data points respectively. The shaded region for high lag time ($\vartheta>5\ s$) shows the range of 5 bootstraps statistics with $N\approx0.3\times10^6$ data points. ({\bf b}) Estimated irreversibility using lower order KLD ($\sigma_2,\sigma_3,\sigma_4$) for the high activity treatment ($N\approx3\times10^6$). The shaded region shows the range of 5 bootstraps statistics with $N\approx1\times10^6$. ({\bf c}) Estimated irreversibility $\sigma$ as a function of high lag time $\vartheta$ for high, medium and low activity. Each line corresponds to a distinct oocyte, with $N\approx1\times10^6$. The error bars show the range of 5 bootstraps statistics with $N\approx0.3\times10^6$. All estimated irreversibility $\sigma$ is calculated using the $k$-nearest neighbor method with $k = 1$ on steady state oocytes.}
	\label{fig2}
\end{figure*}

Our main result, depicted in Fig.~\ref{fig2}a, is an estimate of the irreversibility $\sigma(\vartheta)$ as a function of lag time $\vartheta$ for the high and medium activity steady states, where the effects are most pronounced.  
Strikingly, the estimated irreversibility is a nonmonotonic function of lag time peaked between $\vartheta^*\approx1-3\ s$ for the high activity state and $\vartheta^*\approx8-10\ s$ for the medium activity state, when using the most converged estimate with $k=1$, where $k$ denotes the number of nearest neighbors used in our estimation of the KLD (SI Sec.~1.5, Fig.~S5).
We also verified that the results for the medium activity state are qualitatively similar for $k=2$ and $5$, but smaller in magnitude and noisier (Fig.~S6).

We can understand this nonmonotonic behavior as follows.
For large lag times ($\vartheta\gtrsim 10\ s$), we have coarse grained out too much information, which generically leads to a smaller KLD with smaller irreversibility estimates.
For shorter times ($\vartheta\lesssim 1\ s$), thermal noise dominates our estimates, leading to poorer convergence and a smaller estimated irreversibility.
This is consistent with the statistics of displacements being dominated by the high frequency fluctuations coming from the background reversible thermal noise~\cite{mizuno2007nonequilibrium,fakhri2014high,guo2014probing}.
The irreversible noise acting on our probes originates in the jostling of the myosin motors whose effects on the probes requires a longer time to manifest~\cite{fakhri2014high}.
Thus, the nonmonotonic behavior of the irreversibility (KLD) tells a story that with the smallest lag times, which include all the timescales, the largest dominant contribution is the reversible thermal motion.
Then, as we progressively remove timescales (by increasing the time lag), the thermal fluctuations are smeared over, leading to the emergence of the irreversible active noise, which is eventually coarse grained away resulting in smaller observed irreversibility.
The peak represents a trade-off between the timescale at which irreversibility is most prominent and the suppression due to coarse graining.
The consequence is that the peak reveals the timescale of nonequilibrium irreversibility originating from the myosin on/off rate, which remarkably is consistent with the position-increment autocorrelation cross-over time estimated from Fig.~\ref{fig1}.
However here, we inferred this timescale using a model-independent irreversibility metric, allowing us to characterize the dissipation.

Remarkably, we found that even the observed irreversibility estimated from short time correlations using low-order KLDs $\sigma_n$ for $n=2,3,4$ are non-monotonic with a peak that reveals the correct order of magnitude for the timescale of non-equilibrium activity (Fig.~\ref{fig2}b). Generically, such low-order KLDs provide a looser bound on the actual dissipation (entropy production) of a process. However, for biological experiments where statistics could be limited, this result implies that the irreversibility metric estimated using low-order KLDs $\sigma_n$ can still be useful in providing an estimate of the nonequilibrium timescale in the system.

Beyond the dissipation timescale $\vartheta\gtrsim 8\ s$, we can be confident our estimator is well converged and consistently capturing the detectable irreversibility of the process.
This allows us to infer relative irreversibility of systems with different amounts of nonequilibrium activity.
This is verified in Fig.~\ref{fig2}c, where we have plotted the estimated irreversibility as a function of time lag for three steady states -- high, medium, and low activity -- each for multiple oocytes.
Though the estimates all decrease with lag time, they clearly appear clustered into three zones corresponding to the three activity levels.
The finite-length trajectories results in small statistical error in our estimates (shaded lines) and the dominate variability comes from two biological replicates (Fig.~\ref{fig2}c.) Our observations that irreversibility can distinguish levels of nonequilibrium activity aligns with the observed variation in irreversibility measured in mechanosensory hair cells for different activity treatments~\cite{roldan2018arrow}. Despite the KLD only offering a bound to the physical dissipation, there is a monotonic relationship between the detectable irreversibility and the expected nonequilibrium activity. This supports the hypothesis that cellular activity and irreversibility are linked and that irreversibility can be considered a readout of cellular activity.

\begin{figure*}[ht]
	\includegraphics[width=\textwidth]{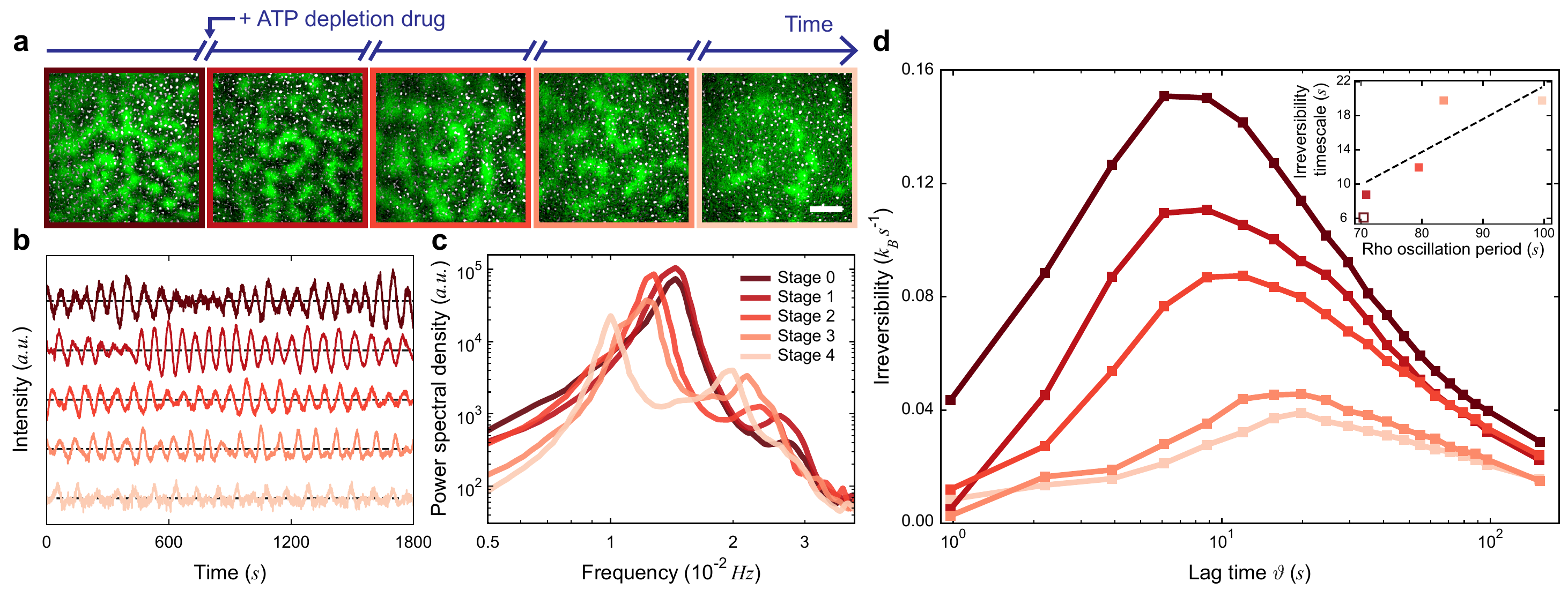}
	\caption{{\bf Time irreversibility reveals the dissipative timescales and levels of non-equilibrium activity in patterned cortex} ({\bf a}) Patterns of Rho signaling waves in Ect2 overexpressed oocytes (green) overlaid on cortical granules (white) before (Stage 0) and after (Stages 1-4) ATP depletion drug is added. Scale bar: $25\ \mu m$ ({\bf b}) Time series of local Rho-GTP concentration oscillations from Stage 0 to Stage 4. ({\bf c}) Power spectral density of local Rho oscillations show a prominent peak at the oscillation frequency. The standard error of the mean is smaller than the linewidths. ({\bf d}) Estimated irreversibility $\sigma$ for the 5 different stages as a function of lag time $\vartheta$. Inset shows the timescale of peak irreversibility as a function of Rho oscillation period. The standard error of the mean for Rho oscillation periods in the inset is smaller than the marker. ($N\approx1.2\times10^5$ for Stage 0, $N\approx3.5\times10^4$ for Stage 1, $N\approx1.6\times10^5$ for Stage 2, $N\approx4.0\times10^4$ for Stage 3, $N\approx1.0\times10^5$ for Stage 4.)}
	\label{fig3}
\end{figure*}

Actomyosin networks are orchestrated in space and time via biochemical signaling proteins. In particular, Rho-GTP is a highly conserved signaling protein pivotal in regulating cellular mechanics. The question we next ask is whether Rho-GTP patterns can tune dissipation timescales? We induced steady state patterns of Rho-GTP by increasing the GEF (guanine exchange factor) responsible for activating Rho-GTP. Rho-GTP patterns maintained constant oscillatory periods within the observed non-equilibrium steady states~\cite{bement2015activator,tan2020topological} (Fig.~\ref{fig3}, see Method). The locally oscillating Rho-GTP pattern (Fig.~\ref{fig3}a) activates myosin molecular motors and modulates actin network organization~\cite{kelkar2020mechanics}, which results in an effective active drive on the cortical granules. The precise activity timescales induced by the Rho-GTP signaling is difficult to ascertain, due to the nonlinear effects in the actomyosin network organization~\cite{kelkar2020mechanics}, but we can vary the Rho-GTP oscillation frequency by modulating the intracellular ATP concentrations via sodium azide treatment (Fig.~\ref{fig3}b-c). Remarkably, the estimated irreversibility calculated from the dynamics of the granules under this perturbation exhibits a shift in the peak timescales in a manner that correlate with Rho-GTP oscillation frequency (Fig.~\ref{fig3}d). Moreover, the magnitude of the irreversibility peak also reflect the changes in intracellular ATP levels. This further corroborates the observations in Fig.~\ref{fig2}c that the observed variation in irreversibility corresponds to physical variations in dissipation.

\subsection*{Analytic validation of irreversibility metric }
Having observed structure in the observed irreversibility of an active biological system, we now turn to validating this approach by analyzing a simple model for a diffusing particle experiencing active noise, such as the cortical granules.  
Consider a two-dimensional overdamped Brownian particle in a viscous fluid with viscosity $\gamma$ at temperature $T$ perturbed by an active noise $\boldsymbol{\mathcal A}(t)$.
The particle's position ${\bf r}_t=(x_t,y_t)$ at time $t$, then evolves according to the Langevin equation
\begin{equation}
\gamma {\dot {\bf r}}(t)=\boldsymbol{\mathcal A}(t) + \sqrt{2\gamma k_{\rm B}T}{\boldsymbol\eta}(t).
\end{equation}
where $\boldsymbol\eta(t)$ is zero-mean Gaussian white noise $\langle \boldsymbol\eta(t)\boldsymbol\eta(s)\rangle={\hat I}\delta(t-s)$.
To make the model tractable, the active noise is taken as a nonequilibrium Ornstein-Uhlenbeck process,
\begin{equation}
   \dot{\boldsymbol {\mathcal A}}(t)=-{\hat K}\boldsymbol{\mathcal A}(t)+\sqrt{2 k_{\rm B}T_A}{\boldsymbol\xi}(t),\quad {\hat K}=\left(\begin{array}{cc}\kappa & \alpha \\-\alpha & \kappa\end{array}\right),
\end{equation}
with zero-mean Gaussian white noise $\boldsymbol\xi(t)$ at active-noise temperature $T_A$.
If ${\hat K}$ were symmetric (${\hat K}={\hat K}^T$), then these dynamics would describe an equilibrium system, as that force on the active nose (${\hat K}\boldsymbol{\mathcal A}$) would be derivable from a potential.
Thus, the activity arises solely due to the asymmetric, off-diagonal elements of the force matrix $\alpha$, which set the timescale of this nonequilibrium mechanism.

\begin{figure*}[ht]
	\includegraphics[width=\textwidth]{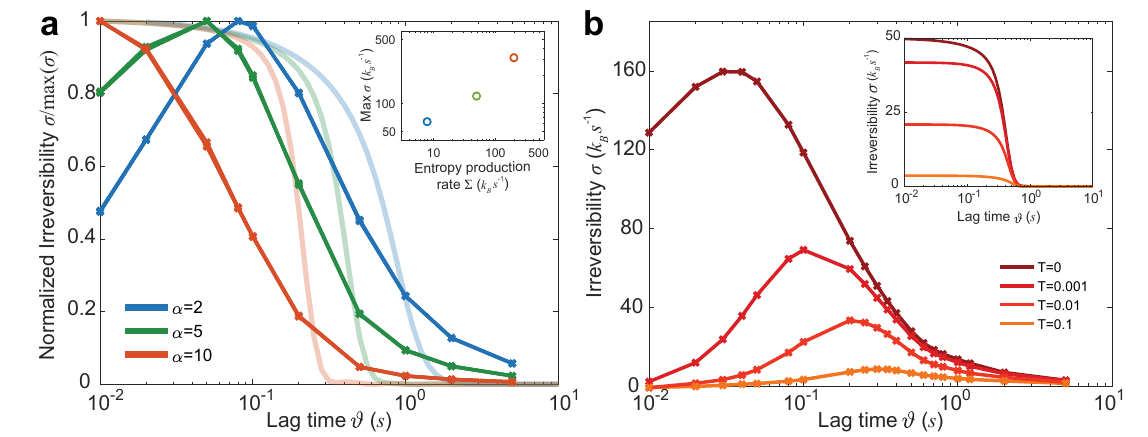}
	\caption{{\bf Time irreversibility reveals the dissipative timescale of a 2D Brownian particle driven by active Ornstein-Uhlenbeck (OU) noise} ({\bf a}) The analytic solution (shaded line) and normalized irreversibility $\sigma/\max(\sigma)$ (solid line) as calculated using our method for $\alpha=2,5,10$ with thermal temperature $T=0$. The inset shows the magnitude of $\sigma$ versus the actual entropy production rate $\Sigma$. ({\bf b}) Irreversibility $\sigma$ as a function of lag time for different thermal temperature $T$ as calculated by our method. Inset shows the corresponding semi-analytic result.}
	\label{fig4}
\end{figure*}

As for the granules, we probe the dissipative timescale $\alpha$ by calculating the  irreversibility $\sigma(\vartheta)$ of the position increments $\Delta {\bf r}^\vartheta(t)={\bf r}(t+\vartheta)-{\bf r}(t)$.
The results of the semi-analytic calculation are presented in Fig.~\ref{fig4}a for three representative values of the (dimensionless) dissipative timescale $\alpha/\kappa = 2, 5, 10$ (low, medium, and high activity).
We observe that there is a pronounced drop-off in the irreversibility controlled by the characteristic scale $\vartheta\approx \kappa/\alpha$, consistent with the experimental observations.
Noticeably, when the coarse-graining exceeds the dissipative timescale ($\vartheta>\alpha$), the irreversibility cannot detect any activity, as expected.

To verify our estimator can detect the dissipative timescale, we simulated trajectories of our model and calculated the irreversibility using the $k$-nearest-neighbor estimator. In the absence of thermal noise ($T=0$), the observed irreversibility of the simulated trajectories at different coarse-graining times $\sigma(\vartheta)$ exhibit a pronounced drop-off at the characteristic timescale $\kappa/\alpha$, in good agreement with the semi-analytic result (Fig.~\ref{fig4}a). Importantly, the maximum value of the observed irreversibility scales with the entropy production of the process (Fig.~\ref{fig4}a, inset), indicating that the magnitude of the observed irreversibility contains information about the relative magnitude of the dissipation, similar to what was observed in the experimental data in Fig.~\ref{fig2}b and ~\ref{fig3}d. 

For non-zero thermal temperature ($T>0$), we find that the observed irreversibility at lower lag time drops to smaller values, revealing a peak near the characteristic nonequilibrium timescale (Fig.~\ref{fig4}b). At timescales shorter than the characteristic timescale, the semi-analytic calculation predicts a progressively smaller (but constant) irreversibility when thermal noise is increased (Fig.~\ref{fig4}b, inset). This numerical result corroborates with the interpretation of our experimental results in Fig.~\ref{fig2}a, whereby at short lag time, the observed irreversibility is dominated by reversible thermal noise, leading to poor convergence and thus small irreversibility.

Nonequilibrium processes are inherently asymmetric under time reversal. Our results demonstrate how a fundamental thermodynamic connection between the physical energy dissipation sustaining a nonequilibrium system and a measure of statistical irreversibility, can provide quantitative insight into the mechanisms of nonequilibrium activity. The dependence of irreversibility on observation frequency (lag time) reveals the striking signature of the timescale of the nonequilibrium activity in a peak of our model-independent irreversibility metric. 
The basic idea of estimating irreversibility for various levels of coarse graining can lead to important inferences whenever there is a well-defined notion of dissipative scales. Analogous procedures could be used to identify spatial scales in living and non-living active matter systems, allowing us to probe the nonequilibrium coupling of structure with function~\cite{skinner2021estimating,guo2021play,seif2020machine,martiniani2019quantifying}.

\medskip
\printbibliography

\section*{Acknowledgments}
This research was supported by a Gordon and Betty Moore Foundation (to N.F.), Sloan Research Fellowship (to N.F.), and the Human Frontier Science Program Career Development Award (to N.F.). We gratefully acknowledge the Gordon and Betty Moore Foundation for supporting T.R.G. and J.M.H. as Physics of Living Systems Fellows through Grant GBMF4513. This research was supported in part through the computational resources and staff contributions provided for the Quest high performance computing facility at Northwestern University which is jointly supported by the Office of the Provost, the Office for Research, and Northwestern University Information Technology. This research was supported in part by the National Science Foundation under Grant No. NSF PHY-1748958. N.F. and J.H. acknowledge discussions during KITP program \lq ACTIVE20: Symmetry, Thermodynamics and Topology in Active Matter\rq.

\section*{Author contributions} 
N.F. and J.M.H. designed and supervised the research. T.H.T. and Y.C.C. performed the experiments. T.H.T., G.A.W., Y.C.C., J.L. and T.R.G. analyzed the data. All authors discussed the results and wrote the paper. 

\section*{Competing interests} 
The authors declare no competing interests.

\end{document}


\maketitle

\section{Materials and Methods}

\subsection{Sample preparation}
\noindent \textbf{Starfish oocytes handling:} Starfish \textit{Patiria Miniata} were procured from South Coast Bio-Marine LLC. The animals were kept in salt water fish tank maintained at $15\:^\circ$C. To collect oocytes, we first made a small incision at the bottom of the starfish to extract some ovaries. The oocytes were then released by carefully fragmenting the ovaries using a pair of scissors. Extracted oocytes were washed twice with calcium free seawater and incubated in filtered seawater (FSW) at $15\:^\circ$C. Experiments were performed within 2-3 days of oocyte extraction. For imaging, oocytes were sandwiched between a coverslip and a glass slide separated by two 60 $\mu$m double stick tape (3M), and sealed with VALAP (1:1:1 vaseline:lanolin:paraffin).\\

\noindent \textbf{Drug treatments:} The mitochrondrial inhibitor drug sodium azide NaN$_3$ (Sigma 71289) and the actin depolymerization drug cytochalasin D (Sigma C8273) were dissolved in water and DMSO to a stock concentration of 2.4 M and 10 mM respectively. These drugs were stored at $-20\:^\circ$C and diluted in FSW immediately prior to use. For ATP-depletion experiments, oocytes were incubated in 2.4 mM NaN$_3$ for 1 hour before imaging. For actin depolymerization experiment, oocytes were incubated in 20 $\mu$M for 30 minutes before imaging. \\

\noindent \textbf{\textit{in vitro} synthesis of mRNA and microinjection:} Four constructs were used in this study. The fluorescently-labeled myosin regulatory light chain MRLC-mEGFP construct was a gift from Peter Lenart~\cite{bischof2017}. The filamentous actin (F-actin) marker Lifeact-mCherry was a gift from Cynthia A. Bradham. To excite Rho waves, we overexpressed Ect2-T808A-mCherry construct, which was a gift from George von Dassow. The Ect2-T808A-mCherry construct is desensitized to one particular Cdk1 phosphorylation, which otherwise limits membrane association during the M-phase. To visualize Rho-GTP molecules, we used fluorescently-labelled rhotekin binding domain construct (eGFP-rGBD), which was purchased from Addgene (Addgene 26732)~\cite{benink2005concentric}. For \textit{in vitro} synthesis of mRNA, we first amplified the constructs by bacterial growth overnight. The plasmids were then purified using Miniprep (Qiagen) and linearized using the appropriate restriction enzymes. MRLC-mEGFP and Lifeact-mCherry mRNA were synthesized using the T7 Ultra and SP6 mMessage mMachine transcription kits (Thermo Fisher Scientific) respectively. To express the constructs, we micro-injected the synthesized mRNA into the cytoplasm of the oocytes and incubated them overnight at $15\:^\circ$C. \\

\noindent \textbf{Rho wave experiments:} Rho-GTP proteins were patterned on the membrane of starfish oocyte to control the oscillations of cortical granules. Ect2-T808A-mCherry construct was overexpressed to induce Rho-GTP wave patterns, and eGFP-rGBD construct was expressed to visualize the Rho-GTP wave patterns. Hormone 1-methyl adenine (1-MA) was added to induce oocyte meiosis, and the construct-expressing oocytes were transferred to the imaging chamber. The imaging chamber was made by attaching a gas-permeable polymer coverslip (ibidi sticky-Slides) to a glass coverslip with two parallel lanes of 100$\mu$m-thick double-sided Scotch tapes. The oocyte chamber was immersed in 400 $\mu$L seawater droplet and isolated in a capped 35mm dish (MatTek glass bottom round dish) to prevent evaporation. The chamber had two open ends, and thus allowed material exchange with the seawater droplet. 4.5 hours after meiosis was induced, metabolism inhibitor drug sodium azide was mixed thoroughly with the seawater droplet to deplete the ATP level in starfish oocytes. The Rho-GTP protein patterns and cortical granules were then recorded using our custom imaging setup.

\subsection{Imaging setup}
\noindent \textbf{Custom imaging setup:} Endogenous granules embedded in the actin cortex of the starfish oocyte are intrinsically fluorescent. The cortical granules were excited either by 488 nm laser (50 mW, OBIS, Coherent Inc.) or 561 nm laser (500 mW tunable, Genesis MX, Coherent Inc.). A neutral-density filter (NDC-50C-4M, Thorlabs) was used to adjust the intensity of the 488 nm laser beam. The beam diameter was expanded using two lenses with focal lengths f1 = 40 mm and f2 = 300 mm (Thorlabs). The expanded beam was circularly polarized using a quarter-wave plate (WPQ05M-561, Thorlabs) and then focused into the back aperture of a high-NA (numerical aperture) oil objective (Nikon CFI Plan Apo 60X/NA=1.40). Emitted fluorescent light was collected through the same objective and passed through a custom made dichroic beam splitter (ZT405/488/561/700-800, Chroma), filtered using a $900 - 1100$ nm band-pass filter (FF01-1001/234, Semrock) and focused onto a short-wave IR camera with an InGaAs detector (NirvanaST 640, Princeton Instrument). For steady state experiments, the videos of cortical granules in the cortex were recorded at 10 frames per second, for $12000 - 15000$ frames. 

The Rho-GTP protein patterns were imaged with the identical imaging setup described above. The Rho-GTP patterns were excited by 488 nm laser using the same light path. The emitted fluorescent light passed through the identical dichroic beam splitter, filtered using a single band-pass filter (FF01-531/46-25, Semrock), and eventually focused onto a EMCCD camera (Andor, Oxford Instrument). For the ATP depletion video sets, the Rho-GTP patterns and the cortical granules were imaged simultaneously at 2 frames per second and 4.09 frames per second, respectively. Five 30-minute videos were taken sequentially, with $\leq$ 5 minute breaks to start drug treatment (between stage 0 and stage 1) and to re-adjust focus plane. Due to photobleaching of fluorescent molecules, the average image intensity in field of view decreases linearly over time. To correct for photobleaching, we rescaled the image intensities based on the instantaneous average intensity in field of view. To correct for spatially-varying background intensity, we obtained a static background by projecting all frames in each video into one averaged image, and subtracted the static background from all frames in each video.\\

\noindent \textbf{Confocal imaging setup:} Fluorescence imaging of Lifeact-mCherry was performed on a Zeiss 710 point scanning confocal system. The system includes a Zeiss AxioObserver motorized inverted microscope stand with DIC optics, a motorized XY stage and three photomutiplier detectors. Images were acquired using a Plan-Apochromat 63x/1.40 Oils DIC M27 objective with 561nm laser excitation and the appropriate beam splitter.  

\subsection{Data Acquisition}

\noindent \textbf{Position-increment trajectories:}
Our samples are recorded for a duration of approximately 30 minutes at a rate of $f=10$ frames per second.
However, for the analyses carried for time lag $\vartheta>8$ s, the videos are down-sampled to $f=2$ frames per second to speed the calculations.
Particle trajectories are obtained by tracking the observed cortical granules using the software u-track \cite{jaqaman2008}, yielding an ensemble of two-dimensional particle trajectories ${\bf r}_n^{p}=(x^p_n,y^p_n)$ as a function of time $t=n/f$ with $p$ the particle index.
For each particle trajectory, we define a position increment at time lag $\vartheta$ by taking the difference between every $\alpha=\vartheta f$ observations  $\Delta{\bf r}^p_{n,\alpha}\equiv {\bf r}_{n+\alpha}^{p}-{\bf r}_n^{p}$ using a rolling window (as illustrated in Fig. S2({\bf b})).
Stationarity of the resulting time series was tested using a unit root test, which either confirmed stationarity in almost every instance or rarely was inconclusive, allowing us to concatenate them together.
Finally, assuming over the time scale of our observation the actomyosin network is homogeneous, we concatenate the position-increment trajectories from our all particles into one long super-trajectory $\Delta{\bf r}_{n,\alpha}$ of length $N$  (Fig S2({\bf d})), which is then used for all subsequent analysis. 

For the Rho-GTP driven granules ATP depletion dataset, the Rho-GTP and granules fields are simultaneously imaged for 30 minutes for each ATP depletion stage. We performed particle tracking of the observed cortical granules using python package trackpy (\url{http://soft-matter.github.io/trackpy/v0.5.0/})~\cite{trackpy}. We obtained the two dimensional granules trajectories ${\bf r}_n^{p}=(x^p_n,y^p_n)$ as a function of time $t=n/f$ with $p$ the particle index.
Assuming that the Rho-GTP oscillation dynamics is spatially homogeneous (as indicated by the heterogeneity $q$ index (Fig. S7)), we concatenate all observed granules trajectories together into a single long super-trajectory for each ATP depletion stage.
To reduce trajectory concatenation discontinuity, we concatenate the granules trajectories by matching their instantaneous Rho-GTP oscillation phases, assuming that granules with similar instantaneous local Rho-GTP concentration are driven by the same active forces. We record the instantaneous Rho-GTP concentration along granules spatial trajectories, and obtain a corresponding Rho-GTP concentration oscillation time series for each granules trajectory. To correct for intensity variation of long timescale, we subtract a 300-second moving average from the Rho-GTP time series before calculating the oscillation phases. The oscillation phase of the Rho-GTP concentration time series can be determined by plotting the 2D phase portrait of $I(t+\tau)$ versus $I(t)$. The oscillation phase is then calculated from $\phi(t)=tan^{-1}(I(t+\tau), I(t))$, where the $\tau$ is selected to be approximately a quarter of Rho-GTP local oscillation period~\cite{tan2020topological}. To stitch the granules trajectories by matching the Rho-GTP oscillation phases, we start from a random granule trajectory, find the next granule trajectory with the starting Rho-GTP phase closest to the end Rho-GTP phase of current granule trajectory, and concatenate the granule trajectories. The search process is repeated until all granules trajectories are concatenated into a single long super-trajectory for each ATP depletion stage.

\subsection{Correlation Analysis}

\noindent \textbf{Calculation of position-increment autocorrelation function $C^\vartheta(\Delta t)$:} The normalized position-increment autocorrelation function at time lag $\vartheta=\alpha/f$ is a function of the time lag $\Delta t = m/f$ defined as
\begin{equation}
C^\vartheta(\Delta t)=\frac{\langle \Delta{\bf r}^p_{n+m,\alpha}\cdot\Delta{\bf r}^p_{n,\alpha} \rangle}{\langle ||\Delta{\bf r}^p_{n,\alpha}||^2\rangle}.
\end{equation}
where the angled brackets represent an average is over length of trajectory (time) and number of observed granules.
All correlation functions were calculated using videos downsampled to $f=2$ frames per second.\\

\subsection{Power spectral analysis}

\noindent \textbf{Power spectral density of local Rho-GTP concentration oscillation.}  To measure how Rho-GTP local concentration oscillation periods change with ATP levels, we analyzed the power spectrum of Rho-GTP local oscillation time series, assuming that at each ATP stage, the Rho-GTP local oscillation period was constant. For each stage (ATP level), 1000 random spatial points were selected in the 30-minute video. To reduce the imaging noise, we averaged each spatial point with its 8 $\mu$m square block neighbors. At each fixed spatial block, the concentration of Rho-GTP oscillates in time between around 50 - 100 seconds. To better resolve the Rho-GTP local oscillation period, we de-trended the Rho-GTP oscillation time series by subtracting a 300-second moving average. The power spectral density of the Rho-GTP intensity local oscillation time series were then obtained by maximum entropy spectral analysis. Maximum entropy spectral analysis was performed using python package memspectrum (\url{https://pypi.org/project/memspectrum/}). The order of autogressive model for fitting the time series was determined using Burg's algorithm and Parzen's criterion on autoregressive transfer function (CAT). As we averaged the power spectrum over 1000 random points at each stage, the averaged power spectrum of Rho-GTP oscillation time series showed a distinct peak at the dominant Rho-GTP local oscillation frequency (Fig. 3({\bf c}) . The oscillation frequency is calculated by selecting the frequency of maximum power spectral density for each spatial point, and averaging over the 1000 points for each video (Fig. 3({\bf d} inset). The dominant Rho-GTP local oscillation frequency decreased as ATP was depleted over time.

\subsection{Irreversibility Analysis}
{\bf Overview:} 
The Kullback-Leibler (KLD) divergence between the distribution of the position-increment trajectories $\Delta{\bf r}_{n,\alpha}$ and the time reverse series $\Delta{\bf r}_{N-n,\alpha}$ is a thermodynamic measure of irreversibility.
As described in detail below, to estimate the KLD we use a $k$-nearest-neighbor method developed in~\cite{wang2009}.
To speed convergence, we employ a heuristic observation that the more similar to independent identically distributed (i.i.d.) random variables is the time series, the faster the estimate converges.
Thus, by using a bijective map of the time series, which does not change the KLD, but results in a more i.i.d.\ time series, can result in faster convergence of our estimates.
Such protocols, are generically called `whitening' procedures; we employ two: (i) we fit an autoregressive model to remove correlations~\cite{Roldan2018}, and (ii) a principal component analysis (PCA) transformation to normalize covariance~\cite{wang2009}. \\

\noindent \textbf{Determining irreversibility:}
We begin with a general discussion on how to estimate the irreversibility, before turning to the specifics of the whitening procedure.
We have in mind a time series of measurements, like our position-increment data, generically denoted as ${\bf x}_1^n=(x_1,\dots,x_n)$.
The irreversibility of this time series is then defined as the limit of the growth rate as $n\to\infty$ of the KLD between forward and reversed time series (Eq. 1 main text $\tau=1/f$):
\begin{equation}
\sigma = \lim_{n\to\infty}\frac{k_{\rm B} f}{n} D({\mathcal P}({\bf x}_1^n)||{\mathcal P}({\bf x}_n^1))\equiv \lim_{n\to\infty}\frac{k_{\rm B} f}{n} D_n , 
\end{equation}
where we have introduced the $n$-th order relative entropy $D_n$~\cite{Roldan2012}.
As it turns out, the ratio $D_n/n$ is known to converge rather slowly.
However, the difference in $n$-th order relative entropies 
\begin{equation}
d_{n}=D_{n}-D_{n-1}.
\end{equation}
converges to the same value, but faster~\cite{Roldan2012}.
Thus, our goal will be to accurately estimate the $D_n$ for large $n$ and then extract the rate of growth from $d_n$.
For our estimates, we chose an $n$ that was large enough so that $d_n$ did not vary noticeably with $n$.
We found that $n=29$ was sufficient to meet this criterion (Fig. S5({\bf b})).  \\

\noindent \textbf{Nearest neighbor estimator:}
To estimate the $n$-th order relative entropy, we used a $k$-nearest-neighbor
estimator described in \cite{wang2009}. 
Imagine we have $N$ samples for our time series ${\bf x}_1^{n,(i)}$, $i=[1,N]$, from which we construct $N$ samples of our reverse time series ${\bf x}_n^{1,(i)}$.
The $k$-nearest-neighbor estimate is constructed by comparing two distances for each sample of the forward time series ${\bf x}_1^{n,(i)}$. The first is the Euclidean distance $\rho_i$ to the $k$-th nearest sample (excluding itself ${\bf x}_1^{n,(i)}$) in the set of forward time series.  The second is the distance $\delta_i$ to the $k$-th nearest sample in the set of reversed time series (excluding its time-reversed self ${\bf x}_n^{1,(i)}$).
This is illustrated in Fig.~S5({\bf a}) for different
choices of $k$. 
The relative-entropy estimator $\hat{D}_{n}$ then takes
the form
\begin{equation}
\hat{D}_{n}= \frac{2n}{N}\sum_{i=1}^N \log{\frac{\delta_i}{\rho_i}}.
\end{equation} \\
 Again, we verified that this estimator plateaued with number of samples.
 For our experiment, we found that for large time lags $\vartheta = 8,15,22,30$~s, $N = 912\,150$ samples were sufficient (Fig. S5({\bf c})). 
 For small time lags $\vartheta = 0.1,0.5,1,2,3,4$~s, $N = 3\,100\,000$ samples were used (Fig. S5({\bf c})) due to worse convergence with smaller
$\vartheta$ values. \\

\noindent \textbf{Position-increment whitening:}
Following~\cite{Roldan2018}, we first whitened the position-increment trajectories by fitting a bivariant autoregressive model of order 1, AR(1), to 
the position-increment time series concatenated with its time 
reverse sequence $\Delta{\bf r}_\alpha=\{\Delta{\bf r}_{n,\alpha} ,\Delta{\bf r}_{N-n,\alpha}\}$. 
(The AR order was chosen using the MATLAB ``arfit'' 
package developed by Schneider and Neumaier \cite{schneider2001}, which utilizes Schwarz's Bayesian Criterion.).
The AR(1) model treats
the data as linearly dependent on the previous time step with parameters $\{c_0,c_1\}$ perturbed by  Gaussian white noise $\eta_i$:
\begin{equation}\label{Eq:AR1}
\Delta{\bf r}_{i,\alpha} \approx c_1\Delta{\bf r}_{(i-1),\alpha}+c_0+\eta_i,
\end{equation}
for all $i=[ 0, 2N ]$.
From the fit, we extract the residuals
\begin{equation}
\xi'_{i,\alpha} = \Delta{\bf r}_{i,\alpha} - c_1\Delta{\bf r}_{(i-1),\alpha}-c_0.
\end{equation} 
Roughly speaking, the better the fit, the more i.i.d.\ are the residuals (cf.~\eqref{Eq:AR1}); thereby, whitening the process.
The effect of this whitening is illustrated in Fig.~S3 with the post-whitening two-point position-increment distribution becoming rounder than the pre-whitened distribution.\\

\noindent \textbf{Subtrajectory ensemble whitening:}
From the residuals $\xi_{i,\alpha}'$, subsequences of length $n$, were constructed and used in the subsequent estimation of the $n$-th order relative entropy $D_n$.
However, before estimating $D_n$, one further whitening procedure~\cite{wang2009} was implemented on the subsequences.
This whitening procedure is a form of principle component analysis (PCA), where we first compute the sample covariance ${\hat C}$ of the AR-whitened forward-reverse concatenated trajectories.
We use this estimate to transform the subsequence $\xi_{i,\alpha}'$ to a new subsequence $\xi_{i,\alpha}$ with a covariance that is nearly the identity.
We have dropped the prime as these twice
whitened observables are now the input into the 
$k$-nearest neighbor estimator for relative entropy. The
effect of this second whitening step is illustrated in Fig.~S4({\bf C}).

\subsection{Theoretical Model}

{\bf Irreversibility of a Brownian particle with active noise:}
Inspired by the motion of the cortical granules, we investigate the irreversibility of an analytically-tractable toy model. 
Consider a two-dimensional overdamped Brownian particle in a viscous fluid with viscosity $\gamma$ at temperature $T$ perturbed by an active noise $\boldsymbol{\mathcal A}(t)$.
The particle's position ${\bf r}_t=(x_t,y_t)$ at time $t$, then evolves according to the Langevin equation
\begin{equation}\label{eq:rDot}
\gamma {\dot {\bf r}}(t)=\boldsymbol{\mathcal A}(t) + \sqrt{2\gamma k_{\rm B}T}{\boldsymbol\eta}(t).
\end{equation}
where $\boldsymbol\eta(t)$ is zero-mean Gaussian white noise $\langle \boldsymbol\eta(t)\boldsymbol\eta(s)\rangle={\hat I}\delta(t-s)$.
To make the model tractable, the active noise is taken as a nonequilibrium Ornstein-Uhlenbeck process,
\begin{equation}\label{eq:Adot}
     \dot{\boldsymbol{\mathcal A}}(t)=-{\hat K}\boldsymbol{\mathcal A}(t)+\sqrt{2 k_{\rm B}T_A}{\boldsymbol\xi}(t),
\end{equation}
with zero-mean Gaussian white noise $\boldsymbol\xi_t$ at active-noise temperature $T_A$ and force matrix 
\begin{equation}
{\hat K}=\left(\begin{array}{cc}\kappa & \alpha \\-\alpha & \kappa\end{array}\right).
\end{equation}
Notice that if $K$ were symmetric (${\hat K}={\hat K}^T$), then this would describe an equilibrium system.
Consequently, the nonequilibrium dissipation manifests in the asymmetric part of ${\hat K}$,
\begin{equation}
    {\hat K}^A=\frac{{\hat K}-{\hat K}^T}{2}=\left(\begin{array}{cc}0 & \alpha \\-\alpha & 0\end{array}\right),
\end{equation}
which solely depends on the nonconservative force $\alpha$ that drives the system out of equilibrium.
The parameter $\alpha$ thus sets the time-scale for this mechanism.
Like for the granules, we probe the dissipative time-scale $\alpha$ by calculating the  irreversibility $\sigma(\vartheta)$ as a function of a coarse-graining time scale $\vartheta$.
To this end, we imagine stroboscopically observing the particle's position in time increments of $\vartheta$, obtaining observations ${\bf r}(n\vartheta)$.
Since the positions ${\bf r}$ are not stationary, we instead consider the stationary position increments $\Delta {\bf r}^\vartheta_n={\bf r}(n\vartheta)-{\bf r}((n-1)\vartheta)$.

As we will see, the theoretical tractability of this model rests on two observations.
First, the increments $\Delta{\bf r}^\vartheta_n$, as a linear transformation of a stochastic Gaussian process ${\bf r}(t)$, are themselves Gaussian.   Second, we rely on the property that for a Gaussian processes information-theoretic quantities can be readily calculated in terms of the processes' power spectrum (Fourier transform of the correlation function)~\cite{Pinsker}.
Let us denote the power spectrum of the position increments as
\[
{\hat S}_\vartheta(\omega)=\sum_{n=-\infty}^\infty e^{ i n \omega}\langle \Delta {\bf r}^\vartheta_n\Delta{\bf r}^\vartheta_0\rangle =\langle |\Delta {\bf r}^\vartheta(\omega)|^2\rangle\qquad (-\pi\le\omega\le\pi),
\]
with its equivalent formulation in terms of the Fourier transform of the increments $\Delta {\bf r}^\vartheta(\omega)=\sum_{n=-\infty}^\infty e^{i n \omega}\Delta{\bf r}^\vartheta_n$. 
For there to be irreversibility, ${\hat S}$ must be manifestly asymmetric with nonzero asymmetric part 
${\hat S}_\vartheta^A(\omega)=[{\hat S}_\vartheta(\omega)-{\hat S}^T_\vartheta(\omega)]/2$. 
Then~\cite{Pinsker}
\begin{equation}\label{eq:irr}
   \sigma(\vartheta)= \frac{1}{2\pi}\int_{-\pi}^\pi d\omega\ {\rm Tr}\left[{\hat S}_\vartheta^{A,T}(\omega){\hat S}_\vartheta^{-1}(\omega)\right].
\end{equation}

The problem now reduces to determining the power spectrum.  
We do this in two steps.
First, we relate the power spectrum of the increments ${\hat S}_\vartheta(\omega)$ to that of the positions, ${\hat S}(\omega)=\langle|{\bf r}(\omega)|^2\rangle$ obtained from the Fourier transform of the positions ${\bf r}(\omega)=\int_{-\infty}^\infty  e^{i \omega t}{\bf r}(t)dt$.
In frequency space, a direct application of the definitions leads to the relationship between the increments and positions
\begin{equation}
    \Delta{\bf r}^\vartheta(\omega)=\frac{1}{\vartheta}(1-e^{ i\omega})\sum_{n=-\infty}^\infty {\bf r}\left((\omega+2\pi n)/\vartheta\right),
\end{equation}
which implies
\begin{equation}\label{eq:powerIncrement}
    {\hat S}_\vartheta(\omega)=\frac{4\sin(\omega/2)}{\vartheta^2}\sum_{n=-\infty}^\infty {\hat S}\left((\omega+2\pi n)/\vartheta\right),
\end{equation}
since $\langle {\bf r}(\omega){\bf r}(\nu)\rangle=\langle |{\bf r}(\omega)|^2\rangle\delta(\omega+\nu)$.
Second, we obtain the power spectrum of the positions by solving the equations of motion [\eqref{eq:rDot} and \eqref{eq:Adot}] in frequency space, leading to
\begin{equation}\label{eq:powerR}
    {\hat S}(\omega)=\langle |{\bf r}(\omega)|^2\rangle=\frac{2k_{\rm B}T_A}{\gamma^2\omega^2}\left(\frac{T}{T_A}+\frac{1}{i\omega +{\hat K}}\cdot\frac{1}{i\omega +{\hat K}^T}\right).
\end{equation}
Now, Eqs.~\eqref{eq:irr}, \eqref{eq:powerIncrement}, and \eqref{eq:powerR}, form a collection of analytic formulas that can be solved in concert to obtain the irreversibility $\sigma(\vartheta)$.
When there is no thermal noise ($T=0$), we can solve these equations exactly:
\begin{equation}
    \sigma(\vartheta)=\frac{4\sin(\alpha\vartheta)^2}{\vartheta(e^{2\kappa\tau}-1)}.
\end{equation}
In the presence of thermal noise, we have been unable to find a closed form solution.  
In this case, we utilized a symbolic mathematics application to numerically  evaluate the formulas.

\clearpage \section{Supplementary Figures}

\noindent \textbf{Figure S1:} To verify that the fluctuations are spatially homogeneous, we performed a spatial local heterogeneity test using the geographical detector q-statistics. We first discretized the field of view into a 5-by-5 grid and calculate the variance of the position increment squared $(\Delta r^\vartheta)^2$ in each of the discretized box. The $q$ index is defined as 
\begin{equation}
q= 1-\frac{1}{N\sigma^2}\sum_{i} N_i \sigma_i^2 
\end{equation} 
A $q$ index of 0 indicates spatial homogeneity, where the sum of fractional variance of the position increment squared is comparable to the global variance. For all our experiments, the q index is less than 0.04 for coarse-grained time up to 15s, indicating that our system is largely spatially homogeneous.
\begin{figure}[h]
	\includegraphics[width=0.6\textwidth]{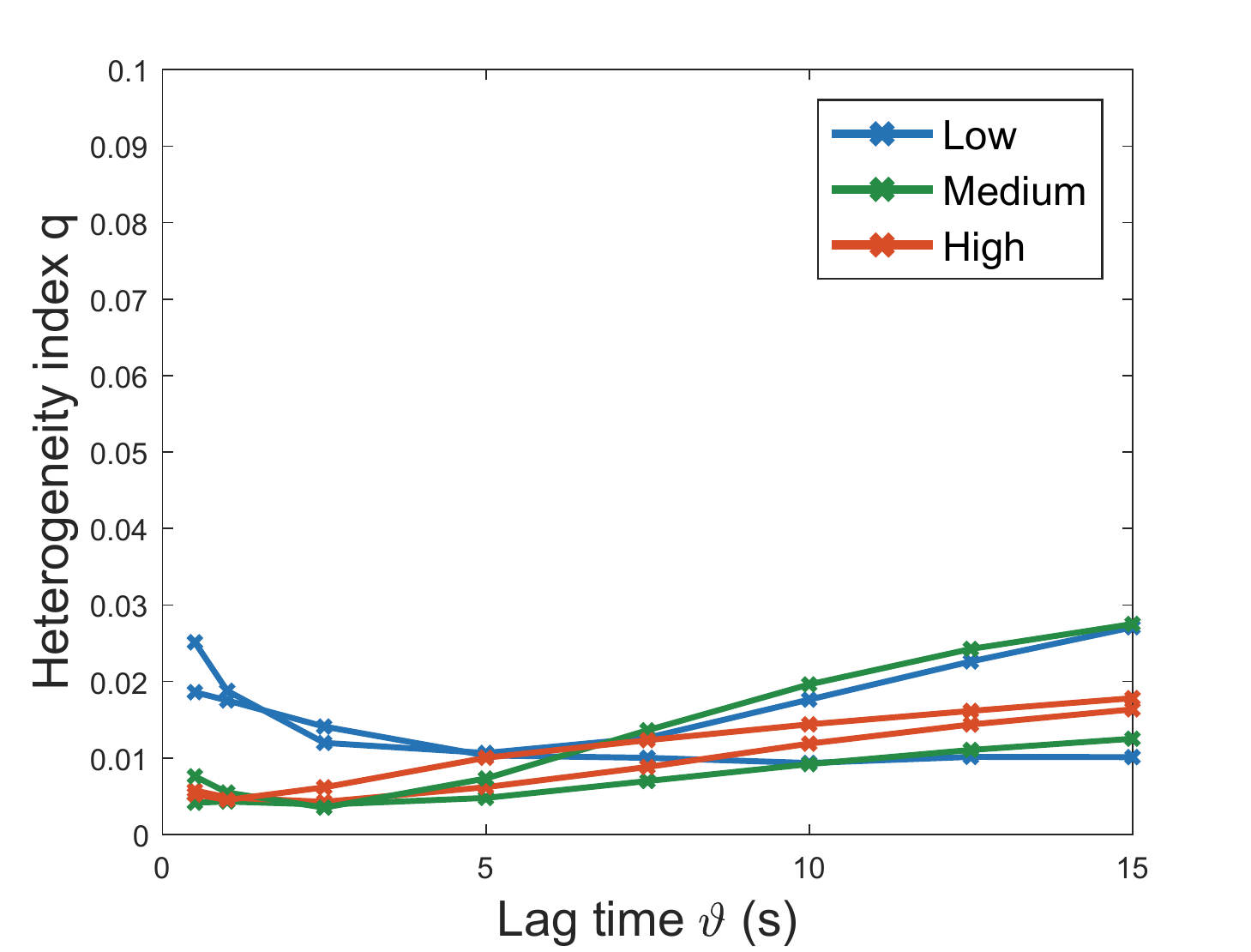}
\end{figure}

\clearpage \noindent \textbf{Figure S2:} ({\bf a}) Sample particle displacement trajectories, ${\bf r}^p(t)$. 
	({\bf b}) Illustration of differencing ${\bf r}_t^\alpha$ into a position-increment series $\Delta {\bf r}^\vartheta(t)$.
	({\bf c}) Observed position increments are stationary time series. ({\bf d}) The position-increment time series $\Delta{\bf r}^\vartheta(t)$
	of the concatenated particle position increments. \\
\begin{figure}[h]
	\includegraphics[width=0.4\textwidth]{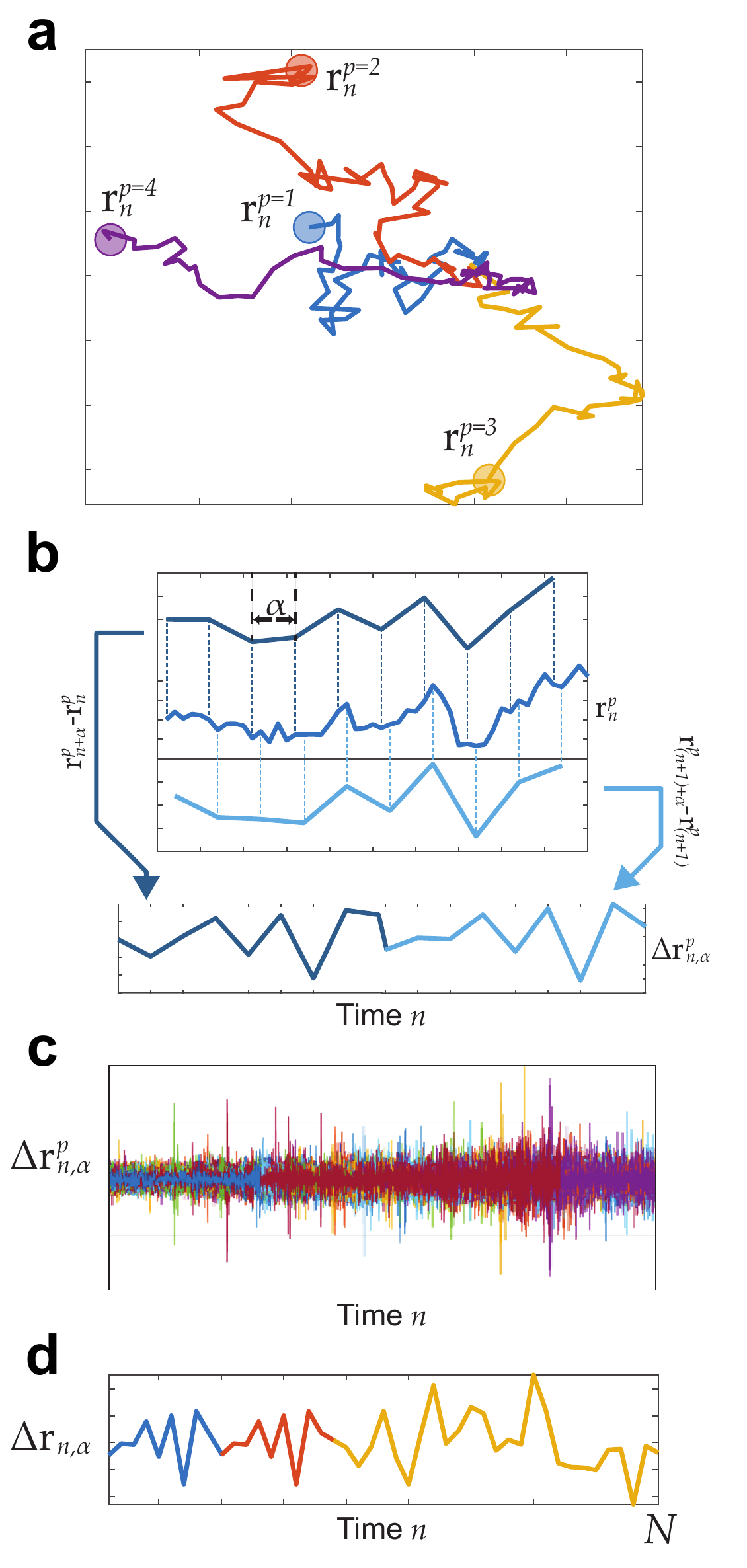}
\end{figure}

\clearpage \noindent \textbf{Figure S3:} Mean squared displacement (MSD) curves for the three different activity levels. For each condition, the solid line is the averaged MSD over more than 300 trajectories while the shaded region indicates the standard deviation. The noise floor is measured by embedding single-walled carbon nanotubes (SWNTs) in agarose gel. The black solid line is the average over three embedded SWNTs while the shaded grey region indicates the standard deviation.
\begin{figure}[h]
	\includegraphics[width=0.6\textwidth]{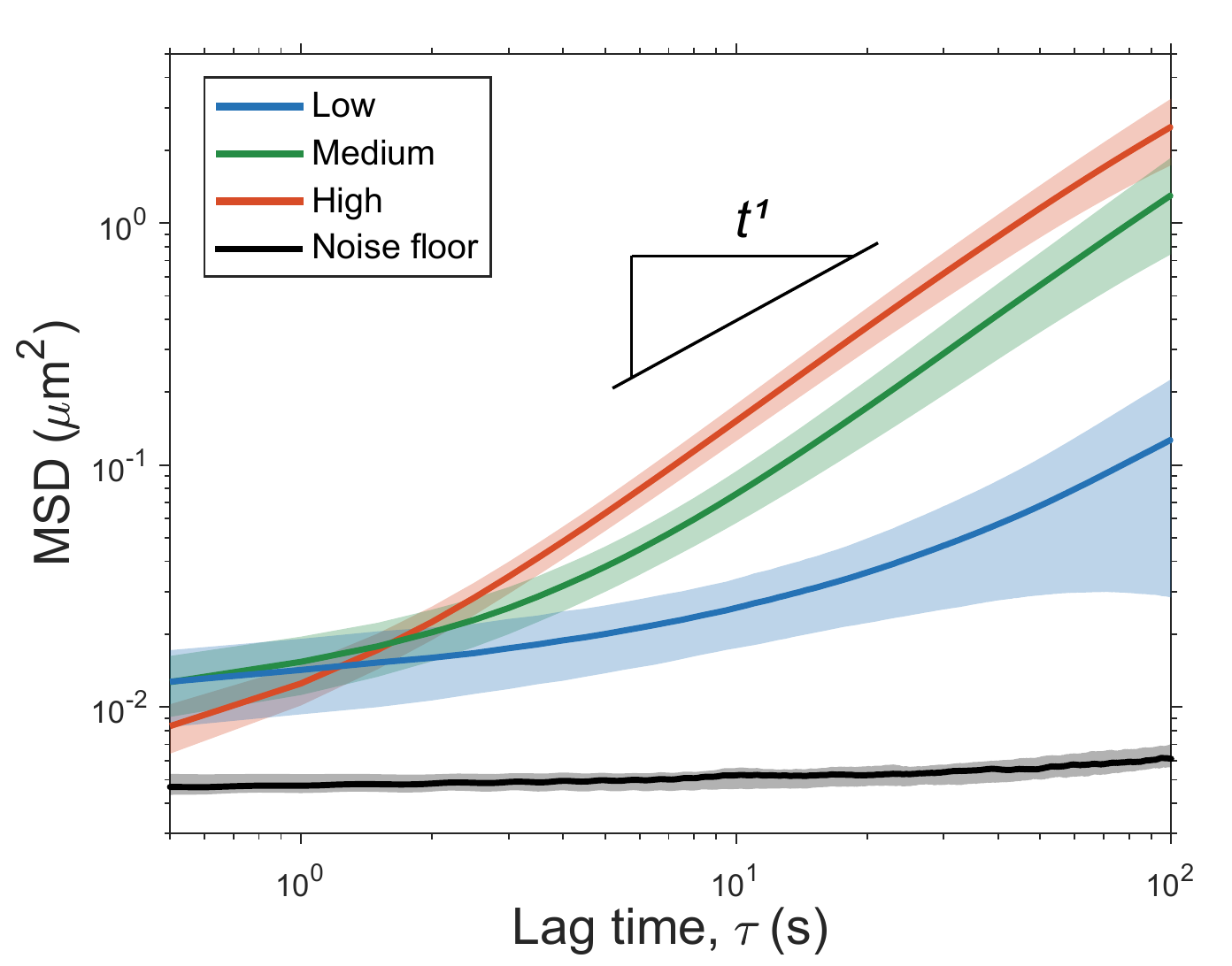}
\end{figure}

\clearpage \noindent \textbf{Figure S4:} Probabilistic correlations between consecutive time steps. ({\bf a}) Two-point probability distribution for the
position-increment trajectories. ({\bf b})  AR(1) whitened
residuals. ({\bf c}) PCA whitened subtrajectories. 
Only $x$-dimension is presented for clarity.
\begin{figure}[h]
	\includegraphics[width=\textwidth]{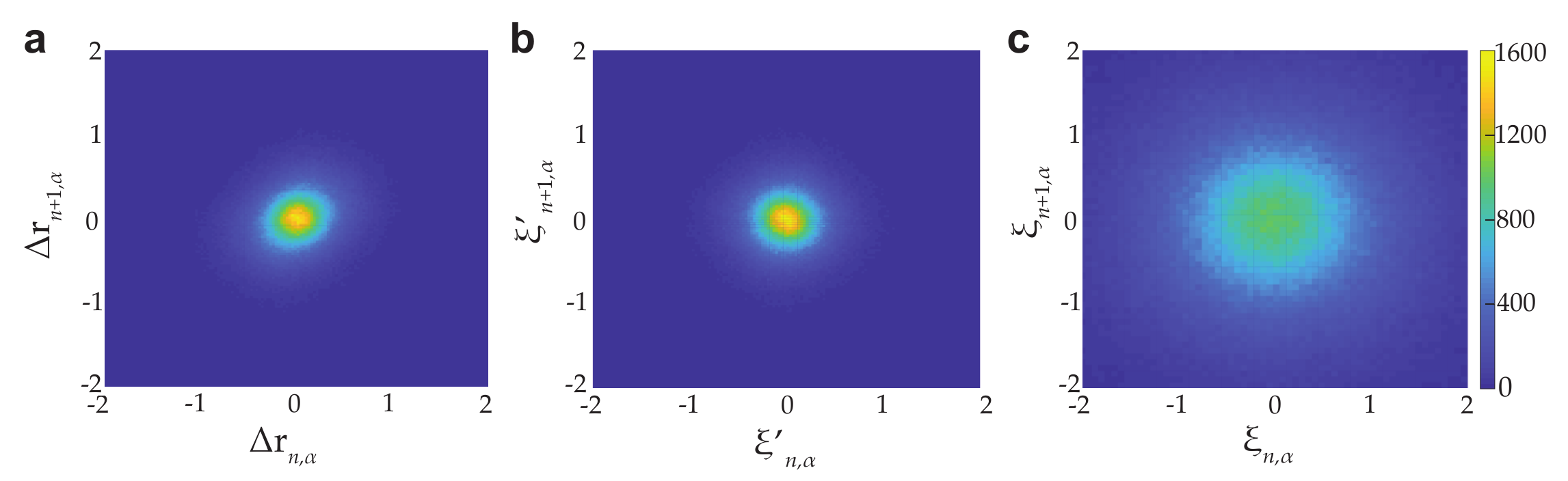}
\end{figure}

\clearpage \noindent \textbf{Figure S5:} ({\bf a}) Demonstration of $k$NN distances $\rho_i$ and $\delta_i$ 
	for $k = 1, 2$. ({\bf b}) Convergence of $d_{29,\vartheta}$ with respect to the length of subsequences $n$ for a wild type oocyte. ({\bf c}) Convergence of $d_{29,\vartheta}$ with respect to the number of samples for the same wild type oocyte. 
\begin{figure}[h]
	\includegraphics[width=1\textwidth]{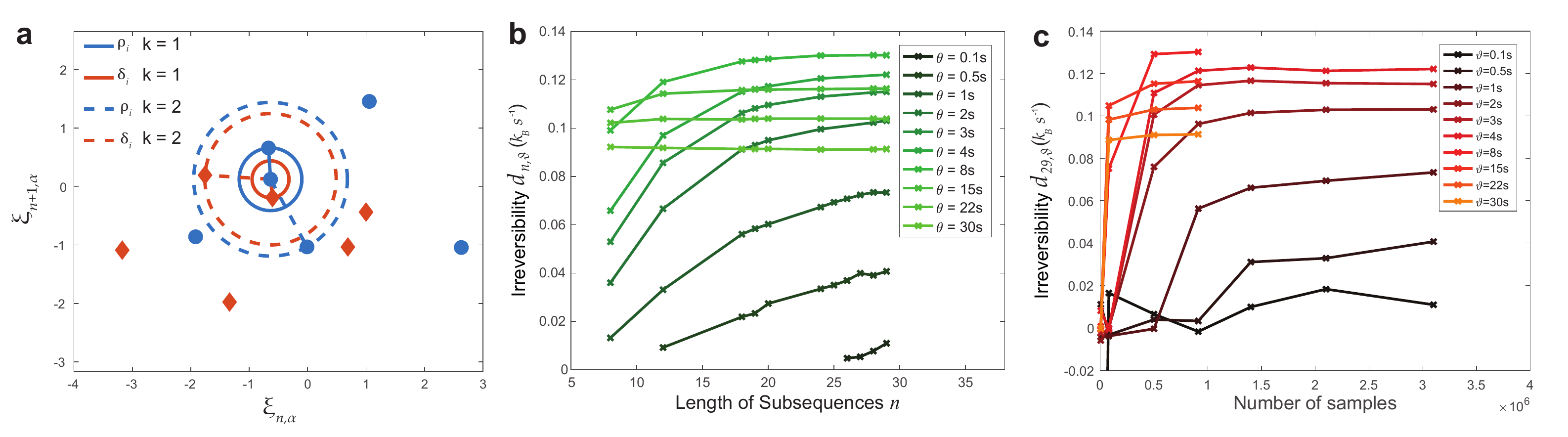}
\end{figure}

\clearpage \noindent \textbf{Figure S6:} Estimated irreversibility $\sigma$ for the medium activity treatment as a function of lag time $\vartheta$ using the $k$-nearest neighbor method for three numbers of nearest neighbors $k=1,2,5$.
\begin{figure}[h]
	\includegraphics[width=0.5\textwidth]{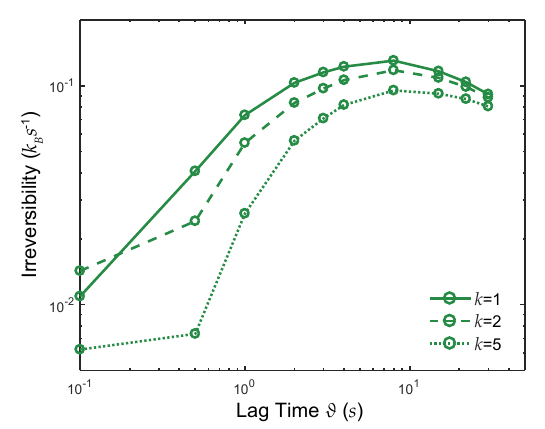}
\end{figure}

\clearpage \noindent \textbf{Figure S7:}  Spatial homogeneity of Rho-GTP local concentration oscillation frequencies at Stage 0-4 ({\bf a-e}). ATP depletion drug was added between Stage 0 and Stage 1. Rho-GTP pattern ATP depletion videos are coarse-grained in space into 4 $\mu$m blocks, and the Rho-GTP concentration local oscillation frequencies are calculated for each block and shown as heatmap. We then discretize the field of view into 5-by-5 grid and calculate the $q$ index for Rho-GTP local oscillation frequencies. The $q$ index is less than 0.03 for all ATP depletion stages. Scale bar, 20 $\mu$m.
\begin{figure}[h]
	\includegraphics[width=\textwidth]{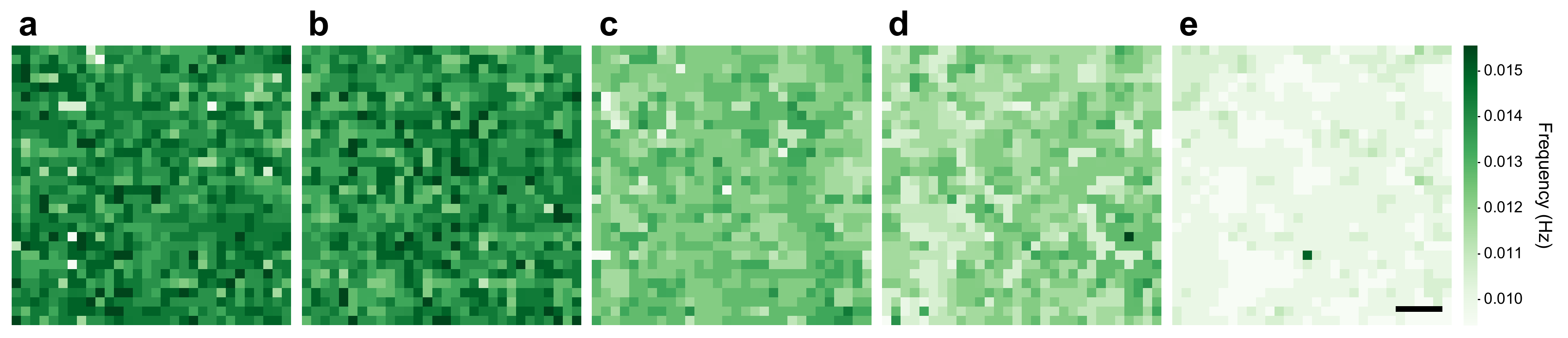}
\end{figure}

\clearpage \noindent \textbf{Figure S8:} ({\bf a}) Convergence of irreversibility with respect to the length of subsequences $n$ for Stage 0 of Rho-GTP wave pattern driven granules trajectories. ({\bf b}) Convergence of irreversibility with respect to the number of samples for Stage 0 Rho-GTP wave pattern driven granules trajectories. Convergence of irreversibility with respect to length or subsequences and to the number of samples are faster for Stage 1 to Stage 4.
\begin{figure}[h]
	\includegraphics[width=\textwidth]{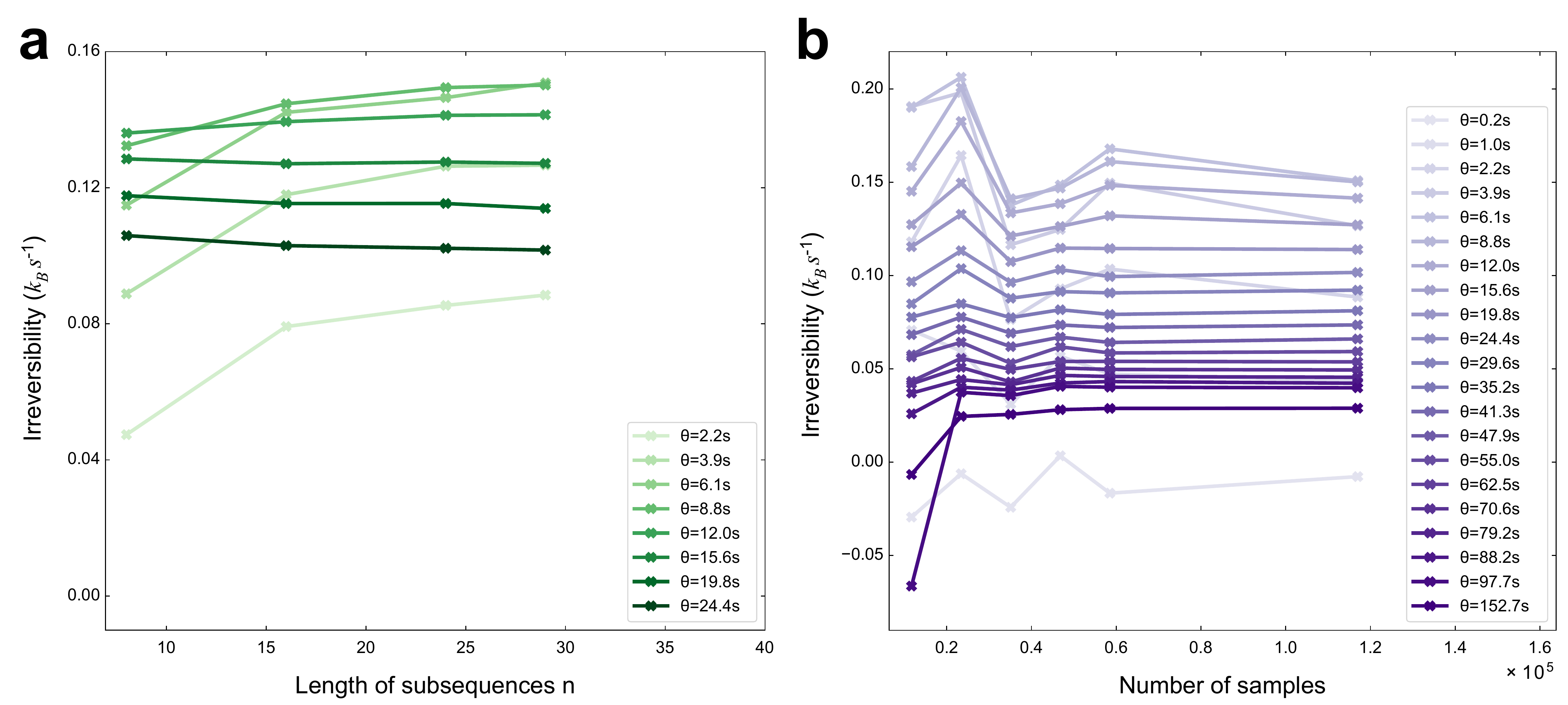}
\end{figure}

\clearpage
\medskip
\printbibliography
